\documentclass[10pt,conference]{IEEEtran}
\IEEEoverridecommandlockouts
\usepackage{cite}
\usepackage{amsmath,amssymb,amsfonts}
\usepackage{algorithmic}
\usepackage{graphicx}
\usepackage{textcomp}
\usepackage{xcolor}
\usepackage{marvosym}
\def\BibTeX{{\rm B\kern-.05em{\sc i\kern-.025em b}\kern-.08em
    T\kern-.1667em\lower.7ex\hbox{E}\kern-.125emX}}
\usepackage{enumitem}
\usepackage{algorithm}
\usepackage{amsmath}
\usepackage{xspace}
\usepackage{listings}
\usepackage[dvipsnames]{xcolor}
\usepackage{subcaption}
\usepackage{colortbl}
\usepackage{tcolorbox}
\usepackage{mdframed}
\usepackage{booktabs}
\usepackage{hyperref}
\usepackage{multirow}
\usepackage{url} 

\newcounter{finding}
\newenvironment{finding}
{
    \refstepcounter{finding}
	\begin{mdframed}[
    	nobreak=true,
    	linecolor=black,
    	roundcorner=12pt,
    	backgroundcolor=gray!05,
    	linewidth=0.5pt,
    	leftmargin=1pt,
    	rightmargin=1pt,
        innerleftmargin=5pt,
        innerrightmargin=5pt,
    	topline=true,
    	bottomline=true,
    	skipabove=10pt
	]
    \textbf{Finding \arabic{finding}}: 
}
{
    \end{mdframed}
    \vspace{3pt}
}

\definecolor{topone}{HTML}{C4483F}
\definecolor{toptwo}{HTML}{EFA9AD}
\definecolor{topthree}{HTML}{F8CACE}
\definecolor{grouphead}{HTML}{F2F4F8}

\makeatletter
\DeclareRobustCommand\onedot{\futurelet\@let@token\@onedot}
\def\@onedot{\ifx\@let@token.\else.\null\fi\xspace}

\newcommand{\rqone}{How are functional abstraction and implementation fidelity related under lightweight probing?}
\newcommand{\rqtwo}{How are functional abstraction and implementation fidelity related in retrieval?}
\newcommand{\rqthree}{Which implementation changes help explain the relationship observed in probing and retrieval?}

\newcommand{\phead}[1]{\vspace{0.25em}{\noindent \textbf{#1}}}

\definecolor{lightgray}{gray}{0.95}

\begin{document}

\title{Lost in the Latent Space: How Code Embeddings Represent Functional Intent and Implementation Fidelity}

\author{
    \IEEEauthorblockN{Lu Sun\IEEEauthorrefmark{1}, Yuheng Huang\IEEEauthorrefmark{1}$^{\textrm{\Letter}}$, Lei Ma\IEEEauthorrefmark{1}\IEEEauthorrefmark{2}, Zhijie Wang\IEEEauthorrefmark{3}}
    \IEEEauthorblockA{
        \IEEEauthorrefmark{1}
        The University of Tokyo, Tokyo, Japan
        \IEEEauthorrefmark{2}
        University of Alberta, Edmonton, AB, Canada\\
        \IEEEauthorrefmark{3}
        Concordia University, Montreal, QC, Canada\\
        \{7779117185, yuhenghuang42\}@g.ecc.u-tokyo.ac.jp,
        ma.lei@acm.org,
        zhijie.wang@concordia.ca}
    \thanks{Yuheng Huang is the corresponding author}
}




\maketitle
\begin{abstract}
    Code embedding models have become a key building block for many software engineering applications, yet their internal representations remain poorly understood. Existing evaluations tend to focus on downstream task performance, offering limited insight into whether a single embedding can support both functional abstraction and implementation-sensitive analysis. To address this gap, we present a large-scale empirical study of 14 state-of-the-art code embedding models. We evaluate their representational behavior through three complementary analyses: supervised probing for functional abstraction and implementation fidelity, zero-shot retrieval experiments to assess geometric similarity, and controlled perturbation experiments that measure embedding responses to semantics-preserving code transformations. The results reveal a consistent abstraction--fidelity tension: models that better preserve stylistic and structural signals tend to do so at the expense of semantic generalization, while models with similar retrieval performance can nevertheless respond very differently to the same implementation changes. Overall, our findings suggest that a single universal code representation remains out of reach, motivating more representation-aware evaluation, selection, and deployment of code embedding models in software engineering.
\end{abstract}






\section{Introduction}



Embedding models\footnote{Throughout this paper, the term embedding model refers to a neural network model trained to produce vector representations of data, rather than the embedding layer of a large language model.} have become a fundamental component of many Software Engineering (SE) approaches for comparing, retrieving, and prioritizing software artifacts at scale~\cite{gu2018deep, zhou2019devign, husain2019codesearchnet, ding2022can, xian2024transformcode, wang2026towards}. By mapping source code, requirements, commits, bug reports, or other artifacts into continuous vector representations, these models enable efficient search and analysis over large software repositories. They have become a core component of many SE tasks, including vulnerability detection~\cite{vuldeepecker, zhou2019devign, wen2025repository}, clone detection~\cite{zhang2019novel}, defect prediction~\cite{hoang2020cc2vec}, code search~\cite{codebert}, and impact analysis~\cite{yan2024enhancing}. In these applications, the quality of the embedding space directly influences which relationships among software artifacts can be identified by downstream classifiers~\cite{vuldeepecker,zhou2019devign,hoang2020cc2vec,zhang2019novel}, ranking algorithms~\cite{husain2019codesearchnet,wang-etal-2025-coderag,tao2025retrieval}, or human analysts~\cite{zeng2023degraphcs, das2023comex, yang2023evacrc}. 

The emergence of Large Language Models (LLMs) did not diminish the need for dedicated embedding models. Although LLMs provide powerful reasoning capabilities over code and natural language, their inference cost, latency, and context limitations make them unsuitable for many downstream or specialized SE tasks that require reasoning over large software repositories or external knowledge sources~\cite{tao2025retrieval,wang-etal-2025-coderag,zhao2026understanding}.
Lightweight embedding models thereby complement LLMs by efficiently searching and retrieval~\cite{gu2018deep,husain2019codesearchnet,wang-etal-2025-coderag,tao2025retrieval}, downstream prediction and ranking tasks~\cite{vuldeepecker,zhou2019devign,hoang2020cc2vec,ding2022can}, and similarity-based analysis~\cite{kang2019assessing, huang2025look, fein2026challenges, Ribeiro2026CodeCorrectness}.
Similar to LLMs, embedding models are also intended to be general-purpose so that they can complement LLM-centric systems and be capable of supporting a diverse range of tasks.
Rather than being optimized for a single downstream task, they aim to learn a representation space in which similar data points are mapped to nearby vectors while dissimilar ones remain distinguishable. Therefore, the same embedding space can support a wide range of downstream applications. From a deployment perspective, reusing a single embedding model across multiple workflows further reduces computational cost and simplifies system design. These practical benefits have been demonstrated by the success of general-purpose embedding models in natural language processing~\cite{neelakantan2022text,wang2022text,bge-m3,muennighoff2023mteb}.
Consequently, recent code-embedding work increasingly embraces this design paradigm by building general-purpose code embedding models, e.g., multi-task retrieval models~\cite{liu2025codexembed,li2025towards}. 
However, we argue that existing evidence is insufficient to determine when this reuse assumption is reliable for SE applications. Previous studies assess code embeddings on individual downstream tasks~\cite{ding2022can,tian2020evaluating,kang2019assessing,fein2026challenges,zhao2024codingptms}, demonstrating task-specific effectiveness while providing only limited insight into whether a single embedding space can simultaneously preserve the diverse forms of software information required across tasks in the LLM era. Intuitively, an embedding is a compressed representation of a software artifact, preserving information useful for one notion of similarity but potentially reducing its ability to represent others. This raises a fundamental question: 

\textit{To what extent can a single code embedding model serve as a reusable building block for software engineering applications, and how do its representational trade-offs constrain that role?}

To answer this core research question, we study code embedding models along two complementary dimensions that capture recurring requirements in SE workflows. The first is \textit{high-level} \emph{functional abstraction}: the extent to which an embedding groups artifacts that implement the same requirement, solve the same problem, or exhibit similar algorithmic behavior despite differences in implementation. The second is \emph{implementation fidelity}: the extent to which a representation faithfully preserves information about how a program is implemented. A high-fidelity representation retains naming conventions, formatting choices, syntactic idioms, structural organization, and other implementation-level regularities. This information is essential for tasks such as predicting compilability from invalid code, identifying implementation conventions, analyzing provenance, and reasoning about project-specific development patterns.

We evaluate these dimensions through three diverse analyses. First, we conduct lightweight probing experiments on frozen embeddings to determine whether functional and implementation-level information can be recovered from the learned representations. Second, we evaluate retrieval behavior directly, since many practical applications rely on the embedding geometry space's neighborhood structure rather than a trained downstream classifier. 
Finally, we conduct a controlled perturbation analysis to explain the geometry behind these differences. We apply semantics-preserving code transformations as counterfactual interventions, each modifying a specific implementation characteristic while preserving the program's intended behavior. By measuring embedding displacements under these controlled transformations, we characterize each model's invariance profile, identifying which implementation variations are treated as irrelevant and which remain geometrically distinguishable. Collectively, these analyses provide a systematic characterization of the trade-offs embodied by modern code embedding models and offer practical guidance for selecting embedding models for different SE applications.

In summary, our contributions are as follows:

\begin{itemize}[leftmargin=*, topsep=0pt]
    \item \textbf{A Representation-Aware Evaluation Framework}: We introduce a systematic evaluation suite comprising four distinct sub-tasks across multiple programming languages. This benchmark is explicitly designed to isolate and evaluate the models through two operational axes: functional intent abstraction and form-level fidelity.

    \item \textbf{A Large-Scale Empirical Study}: We conduct a systematic evaluation of 14 SOTA embedding models, spanning both general-purpose and code-specific approaches. Our analysis provides insight into their representational preferences and abstraction behaviors, and reveals a consistent trade-off between stylistic fidelity and semantic abstraction.

    \item \textbf{A Semantic-Preserving Perturbation Study}: We design an unsupervised perturbation study using functionality-preserving code transformations to examine the models' underlying invariance behavior. By tracking the resulting geometric shifts, we identify which forms of local implementation variation are absorbed by the representation and which are preserved as distinct geometric signals.
\end{itemize}

To support future research, we make our code and datasets, including our independently collected AtCoder corpus, publicly available at \href{https://anonymous.4open.science/r/CodeEmbedBench-816922/}{Anonymous Git}~\cite{codeembedbench2026}. The repository also provides preliminary results on CodeSearchNet~\cite{husain2019codesearchnet} and BigCloneBench~\cite{svajlenko2014towards}, together with additional uncertainty and robustness analyses of the observed correlations.

\section{Background and Related Work}

In this section, we first introduce the backgrounds about embedding models. Then we discuss the related work.

\subsection{Text and Code Embedding}



\subsubsection{Definition}

An embedding model is a parameterized function that projects unstructured input data into a continuous high-dimensional vector space, commonly referred to as a latent space or embedding space. Formally, given an input domain $\mathcal{X}$, such as a vocabulary of discrete tokens, an embedding model defines a mapping:

\begin{equation}
    f_\theta: \mathcal{X} \rightarrow \mathbb{R}^d, 
\end{equation}
where $d$ is the dimensionality of the continuous vector space ($d \in \mathbb{Z}^+$) and $\theta$ represents the learnable model parameters. An embedding model is trained by optimizing $\theta$ to minimize the distance between representations of related inputs while maximizing the distance between representations of unrelated inputs in the embedding space, guided by a task-specific supervision signal (e.g., semantic equivalence between programs). This is commonly formulated using contrastive, ranking, or reconstruction-based objectives~\cite{codebert, PLBART, neelakantan2022text, xian2024transformcode}. 

A downstream DNN classifier $H: \mathcal{X} \rightarrow \mathcal{Y}$ built on top of the embedding model $f_\theta$, can be formally decomposed into the composition of two functions:

\begin{equation}
    H(x) = (g_\phi \circ f_\theta)(x) = g_\phi(f_\theta(x)),
\end{equation}
where $\mathcal{Y}$ represents a discrete label space of $K$ classes.

In this formulation, the embedding model $f_\theta: \mathcal{X} \rightarrow \mathbb{R}^d$ serves as a non-linear feature extractor, mapping the input to a continuous latent space. The function $g_\phi: \mathbb{R}^d \rightarrow \mathcal{Y}$ operates as the tail classifier, mapping the $d$-dimensional embedding to the target label space. In practice, this formulation is the driving force behind many standard DNN-based code classification tasks. Across applications such as software vulnerability detection~\cite{vuldeepecker, zhou2019devign, wen2025repository}, impact analysis~\cite{yan2024enhancing}, just-in-time defect prediction~\cite{hoang2020cc2vec}, and code clone detection~\cite{zhang2019novel}, the pipeline typically follows the same pattern: $f_\theta$ encodes task-relevant code semantics, while a lightweight $g_\phi$, usually a Multi-Layer Perceptron (MLP), produces the final prediction.



\subsubsection{Advancement of general-purpose embedding models}

Beyond task-specific code representation learning, a natural question is whether a general representation model can simultaneously capture how a program behaves and how it is implemented. This question has driven the rapid evolution of general code embedding models.
Early research adapted NLP techniques~\cite{word2vec} to source code, but researchers quickly recognized that Programming Language (PL) possess strict, formal grammatical rules unlike Natural Language (NL). As a result, subsequent work began incorporating structural priors explicitly into representation learning. Models such as Code2vec~\cite{code2vec} and Code2seq~\cite{code2seq} leveraged ASTs by extracting syntactic paths and embedding them into vector representations. Later work, including InferCode~\cite{bui2021infercode}, advanced this direction through self-supervised learning on AST structures, training models to predict subtrees to enable more powerful embedding learning.

The introduction of the Transformer architecture marked a major shift toward large-scale pre-training for code representation~\cite{HedgeCode}. Early encoder-based models such as CodeBERT~\cite{codebert} established a foundation for jointly embedding NL and PL into a shared vector space through masked language modeling. Subsequent models incorporated additional structural signals to better capture program semantics~\cite{liu-etal-2023-code, LongCoder}. Examples of such include GraphCodeBERT~\cite{graphcodebert}, UniXcoder~\cite{unixcoder}, PLBART~\cite{PLBART} and CodeT5~\cite{codet5}.


The paradigm has recently shifted toward unified representation spaces driven by contrastive learning and the immense scale of LLMs. OpenAI's early work~\cite{neelakantan2022text} demonstrated that training large models with contrastive objectives can produce highly expressive, general-purpose embeddings. 
This direction has continued to evolve with the emergence of large-scale code models~\cite{li2023starcoder, roziere2023code}. Building upon these foundations, recent embedding models such as CodeXEmbed~\cite{liu2025codexembed} and CodeR~\cite{li2025towards} represent the current SOTA. 
Our work is orthogonal to the development of these novel architectures and pre-training paradigms. Rather than proposing a new model, we empirically investigate the latent spaces these systems construct, revealing fundamental trade-offs that can guide the design of next-generation training methodologies.



\subsection{Empirical Studies on Embedding Models}
\label{subsec:embedding-evaluation}

Evaluating embeddings as general-purpose, reusable representations requires moving beyond isolated downstream metrics.
In Natural Language Processing (NLP) and adjacent domains, standardized benchmark suites such as MTEB~\cite{muennighoff2023mteb} and MAEB~\cite{assadi2026maeb} assess embedding spaces across diverse task families, reflecting the operational role of embeddings as task-agnostic, reusable representations. 
Rather than treating an embedding model as a specialized, single-purpose component, these benchmarks characterize the structural and semantic relations sustained within the latent space across use cases.

However, different from the benchmark-driven evaluation paradigm adopted in NLP, research on code embeddings in software engineering has largely focused on downstream task performance. 
Prior studies have investigated whether pre-trained code embeddings improve SE-specific prediction tasks~\cite{ding2022can}, assessed their generalizability across distinct domains~\cite{kang2019assessing}, examined how code-change representations predict patch correctness~\cite{tian2020evaluating}, and analyzed the behavioral characteristics of code embeddings in industrial production environments~\cite{fein2026challenges}. 
Recent empirical efforts further explore methodologies for optimizing embedding extraction from pre-trained models and guidelines for model selection in safety-critical contexts, such as vulnerability detection~\cite{zhao2023get, zhao2024codingptms}. 
While these studies conclusively demonstrate that model architecture and embedding configurations heavily influence downstream performance, their evaluations remain predominantly tied to extrinsic task utility. 
As a result, these studies primarily answer which embedding performs best for a particular application, providing comparatively limited insight into the information encoded by the representation space itself.

The growing adoption of LLM-centric systems further raises the importance of understanding embedding spaces beyond classifier-level downstream performance. In many practical workflows, frozen embeddings directly determine which code snippets are retrieved for search, repository-level context selection, or retrieval-augmented generation. This dependence has motivated a growing body of studies that investigate embedding quality in retrieval-oriented settings. CodeSearchNet~\cite{husain2019codesearchnet} established a large-scale benchmark for semantic code retrieval. CoSQA and CoSQA+\cite{huang-etal-2021-cosqa,gong2024cosqa+} extended this perspective to question-driven retrieval, where models must align code with natural-language information needs rather than short search queries. Broader code information retrieval benchmarks~\cite{li2024coircomprehensivebenchmarkcode} further expanded the evaluation scope, while CodeRAG-Bench~\cite{wang-etal-2025-coderag} examined the role of retrieval quality in retrieval-augmented code generation pipelines.

Our work complements these task-centered paradigms. 
Rather than benchmark-hunting to determine which model achieves peak performance on a specific SE task, we systematically investigate the intrinsic properties and types of information that reusable code embedding spaces preserve. 
Specifically, we study code embeddings from two complementary representational perspectives: 

\begin{itemize}[leftmargin=*, topsep=0pt]
    \item \textbf{Functional Abstraction:} The capacity to align functionally equivalent code fragments closely within the latent space, irrespective of variations in syntax, programming language, or implementation mechanics.
    \item \textbf{Implementation Fidelity:} The capacity to preserve structural anomalies, stylistic regularities, and surface-level implementation characteristics crucial for syntax-aware tasks.
\end{itemize}

To rigorously evaluate these dimensions, we employ a multi-faceted methodology consisting of probing tasks to measure the recoverability of features from frozen embeddings, zero-shot retrieval tasks to observe localized geometric tendencies in deployment-like configurations, and controlled perturbation analysis to quantify vector shifts under precise semantic and syntactic code mutations. 
Rather than asking which embedding performs best on a particular benchmark, our goal is to understand what information different embedding spaces preserve, the trade-offs they embody, and whether they are suitable as general-purpose representations for software engineering.

\section{Study Design}

A central challenge in evaluating code embeddings is that there is no single notion of program similarity that serves all software engineering tasks. The same pair of programs may be regarded as similar because they implement the same functionality, or different because they employ distinct implementation strategies.
Consequently, every embedding space implicitly prioritizes certain software properties while abstracting away others. Rather than treating these design choices as hidden assumptions, we make them explicit by asking three complementary questions: What software information remains recoverable from the embedding? Which software properties primarily determine geometric similarity? And which forms of implementation variation are intentionally absorbed by the representation?


In this study, we instantiate these questions through two complementary representational axes. The first is \emph{functional abstraction}. 
A functionally abstract representation should place together implementations that realize the same computational intent despite differences in programming language, syntax, or program structure. The second is \emph{implementation fidelity}. 
A high-fidelity representation should retain naming conventions, formatting choices, syntactic idioms, structural organization, and other implementation-level regularities that may be irrelevant to functional intent but remain valuable for form-sensitive software engineering tasks.


We first examine these axes through probing analysis. Probing has been widely used to study what information is encoded in neural representations by training lightweight predictors on frozen embeddings~\cite{alain2016understanding, hewitt2019designing}. In our setting, probing asks a simple scientific question: \emph{once source code has been compressed into embeddings, which software properties remain recoverable without updating the embedding model?}

We instantiate the functional-abstraction axis through \emph{Algorithm Tag Prediction}. Algorithm tags capture strategy-level concepts that relate implementations beyond their surface form, such as dynamic programming, graph traversal, or greedy algorithms. Although they do not fully characterize program semantics, they provide an observable signal of whether an embedding preserves high-level functional information across diverse implementations. We instantiate the implementation-fidelity axis through \emph{Author Attribution}. Authorship is particularly suitable because it emerges from the aggregation of many subtle implementation-level cues, including naming habits, formatting preferences, code decomposition patterns, idiomatic constructs, and recurring local design choices, rather than any single syntactic feature.

Probing, however, only evaluates whether a learned prediction head can recover a signal from the embedding. In many practical applications, embeddings are used directly through their geometry, where nearest-neighbor relationships determine retrieval, clustering, and ranking behavior. We therefore complement probing with zero-shot retrieval experiments to test whether functional abstraction or implementation fidelity naturally governs geometric similarity. 

Finally, we use controlled perturbation analysis to move beyond task-level performance and examine the representation's underlying invariance behavior. Following the broader interpretability principle of studying model behavior through controlled input changes and their corresponding changes in model responses~\cite{ribeiro2016should,lundberg2017unified}, we apply functionality-preserving code transformations that selectively modify implementation details while maintaining program behavior. We then measure which changes leave the embedding effectively unchanged and which produce substantial geometric shifts. This perspective characterizes the representation's selective invariance~\cite{anselmi2016invariance, lenc2015understanding}: \emph{not merely whether it is robust to code variation, but which distinctions it suppresses, preserves, or amplifies.}


We seek to investigate the following research questions:

\begin{itemize}[leftmargin=*, topsep=0pt]

    \item {\em RQ1}: {\rqone}

    \item {\em RQ2}: {\rqtwo}

    \item {\em RQ3}: {\rqthree}

\end{itemize}


The remainder of this section describes the datasets, benchmark construction, and evaluation protocols used to investigate these research questions.
Table~\ref{tab:benchmark-overview} provides a compact overview of the resulting five evaluations.

\begin{table}[t]
    \caption{Overview of our evaluations.}
    \label{tab:benchmark-overview}
    \footnotesize
    \centering
    \setlength{\tabcolsep}{3.0pt}
    \renewcommand{\arraystretch}{1.08}
    \begin{tabular}{@{}>{\raggedright\arraybackslash}p{0.55cm}>{\raggedright\arraybackslash}p{2.5cm}>{\raggedright\arraybackslash}p{4.35cm}@{}}
        \toprule
        \textbf{RQ}             & \textbf{Evaluation}                 & \textbf{Intent} \\
        \midrule
        RQ1                     & Algorithm-Tag Prediction            & Measure recoverability of functional abstraction in frozen embeddings. \\
        \rowcolor{lightgray}RQ1 & Author Attribution                  & Measure recoverability of implementation fidelity in frozen embeddings. \\
        RQ2                     & Cross-Language Functional Retrieval & Test whether code with the same problem tags is retrieved across languages. \\
        \rowcolor{lightgray}RQ2 & Cross-Task Author Retrieval         & Test whether code by the same author is retrieved. \\
        RQ3                     & Controlled Transformation Analysis  & Measure how controlled implementation changes move embeddings. \\
        \bottomrule
    \end{tabular}
\end{table}


\subsection{Dataset Design}
\label{subsec:dataset}


Our evaluation dataset is built from three public \emph{multilingual} corpora: Open-R1 Codeforces~\cite{penedo2025codeforces}, Google Code Jam (GCJ)~\cite{10.1145/3472410.3472445}, and HumanEval-X~\cite{zheng2023codegeex}. We use Codeforces for algorithm labels and cross-language functional retrieval, GCJ for authorship and author retrieval, and HumanEval-X for controlled transformations. The cleaning steps below define the datasets used by the RQs.

\phead{Codeforces (RQ 1,2)}. Open-R1 Codeforces is a dataset of competitive programming submissions. We use the C++, Java, and Python submissions, which represent the three most popular programming languages in the dataset, while retaining the problem-level algorithm tags. Submissions often mix the core algorithm with parsing, helpers, and problem-specific scaffolding, which can disperse the representational signal away from the target algorithmic evidence. We therefore clean the code within each language before constructing the aligned dataset. After structuring, each retained record keeps the source code, language, problem identity, contest identity, and algorithm tags. The final dataset contains 3,712 problems; for each problem, we keep one implementation per language, yielding 11,136 implementations.

\phead{GCJ (RQ 1,2)}. GCJ is another dataset of competitive programming submissions with user profile information. We first exclude malformed rows and infer programming language from file extensions. Then the remaining records are de-duplicated at the (language, username, task) level so repeated submissions for the same task do not dominate an author's profile. Source code longer than 10,000 characters is also excluded. After structuring, each retained record keeps the source code, language, author identity, and task identity. To keep author-related tasks closed-set and balanced, the dataset retains authors with at least 30 valid implementations and uses the 50 most prolific authors in each language. This gives each retained class enough training examples while preventing a few prolific authors from dominating the evaluation. 
The dataset contains 6,684 C++, 5,078 Java, and 3,231 Python implementations.

\phead{HumanEval-X (RQ3)}. HumanEval-X was designed for evaluating the multilingual ability of code generation models.
We retain 100 original programs per language from the C++, Java, and Python HumanEval-X solutions. Compared with longer competitive-programming submissions, these self-contained functions reduce boilerplate and problem-context noise. Their clean, function-level structure makes HumanEval-X particularly well suited for perturbation analysis, where the goal is to isolate the representational effects of controlled implementation changes

\phead{Controlled Implementation Transformations (RQ3)}.
To study which implementation characteristics are preserved or abstracted away by an embedding model, we construct controlled five categories of code transformations that modify one implementation dimension at a time while preserving program functionality. Table~\ref{tab:rewrite-family-design} lists the details. 

\begin{table}[!t]
    \caption{Controlled implementation transformations.}
    \label{tab:rewrite-family-design}
    \centering
    \footnotesize
    \setlength{\tabcolsep}{3pt}
    \begin{tabular}{@{}p{2.4cm}p{6.0cm}@{}}
        \toprule
        \textbf{Category} & \textbf{Controlled change}                                                                           \\
        \midrule
        Data and Declaration    & Modify literals, initialization patterns, declarations, or value-preserving container forms.        \\
        Control Flow            & Rewrite loops, branches, and condition structure while preserving the executed behavior.             \\
        Function                & Restructure helper boundaries and abstraction structure while preserving the externally tested API.      \\
        Dead/Bogus Code         & Insert semantically inert code such as unreachable branches or unused helpers. \\
        Format \& Rename        & Modify lexical and layout cues such as local identifiers, whitespace, wrapping, and comments.       \\
        \bottomrule
    \end{tabular}
    \vspace{-4mm}
\end{table}

\phead{Edit Magnitude.}
Transformation families specify \emph{what} implementation characteristic is modified, while edit magnitude quantifies \emph{how much} of the implementation changes. To distinguish these two factors, we measure edit magnitude at the logical-block level rather than individual tokens, allowing structural implementation changes to be captured while reducing sensitivity to superficial lexical edits.
Let $B(x)=(b_1,\dots,b_m)$ denote the logical-block sequence extracted from code $x$, where each $b_i$ is a maximal contiguous source block delimited by blank lines, statement-ending tokens, braces, or control-structure prefixes. Let $\mathcal{M}_{x,x'}=\mathcal{M}(B(x),B(x'))$ denote the non-equal opcode tuples returned after aligning $B(x)$ and $B(x')$ with \texttt{SequenceMatcher}~\cite{ratcliff1988gestalt}.
The edit magnitude ratio is defined as:

\begingroup\small
\[
    r(x,x')=\min\!\left(1,\frac{\sum_{(i_1,i_2,j_1,j_2)\in \mathcal{M}_{x,x'}}\max(i_2-i_1,j_2-j_1)}{\max(\lvert B(x)\rvert,\lvert B(x')\rvert,1)}\right).
\]
\endgroup

Based on the edit magnitude ratio, transformations are categorized into three edit-magnitude levels: low ($0\le r\le0.33$), medium ($0.33<r\le0.66$), and high ($r>0.66$). For all transformation families except Format \& Rename, comments are removed before constructing $B(\cdot)$ so that edit magnitude reflects implementation changes rather than comment edits. For Format \& Rename, comments are retained because they constitute part of the intended transformation.

\phead{Generation and Validation.}
We generate rewrites iteratively with GPT-5.4 xhigh under prompts specific to each transformation family. A transformed program is retained only if it satisfies four validation criteria:
\begin{enumerate}
    \item the rewrite ratio falls within the intended strength range;
    \item all 100 black-box tests pass;
    \item Claude Opus 4.5 independently verifies that the requested transformation has been correctly applied while preserving functionality;
    \item sampled rewrites pass manual inspection.
\end{enumerate}
Manual inspection covers 20\% of the retained rewrites. Across the five transformation categories and three strength bands, the retained dataset contains 1,500 test-passing pairs per language and 4,500 pairs in total.

\begin{figure}[t]
    \centering
    \includegraphics[width=\columnwidth]{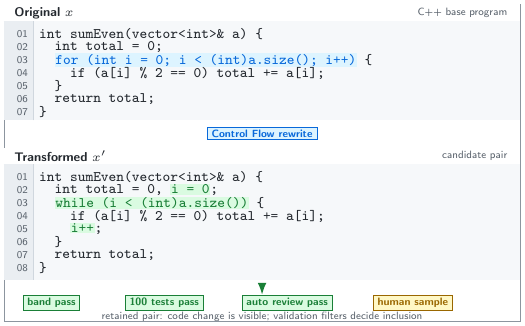}
    \caption{An example of implementation changes from our study. The original program's for-loop was replaced by a while-loop.}
    \label{fig:rq3-change-analysis}
    \vspace{-3mm}
\end{figure}

\subsection{RQ1: Probing Abstraction and Fidelity}

RQ1 is instantiated through two probing tasks. For each embedding model, the encoder is frozen and separate lightweight prediction heads are trained for Algorithm-Tag Prediction and Author Attribution.


\phead{Algorithm-Tag Prediction.}
We evaluate whether algorithmic information remains recoverable from frozen embeddings using multi-label prediction of Codeforces algorithm tags. Separate probes are trained for C++, Java, and Python, and performance is measured by sample-averaged F1.
We train and evaluate a separate probe for C++, Java, and Python and report sample-averaged F1.
\begin{itemize}[leftmargin=*, topsep=0pt]
    \item \textbf{Labels}: The prediction targets are 37-dimensional multi-hot vectors over the tags that remain after filtering rare tags.
    \item \textbf{Split}: The supervised split is 70/10/20 (train/validation/test) by problem. All retained implementations of the same problem stay in the same split, so the test set does not share a problem with the training set.
\end{itemize}

\phead{Author Attribution.}
We evaluate whether author-specific implementation patterns remain recoverable by predicting the author of each submission. Separate probes are trained for each language, and performance is measured by macro-F1.

\begin{itemize}[leftmargin=*, topsep=0pt]
    \item \textbf{Labels}: The prediction target is one author label among the 50 retained authors for each language.
    \item \textbf{Split}: After de-duplication, each author's submissions are split 70/10/20 (train/validation/test), with no retained implementation appearing in more than one split.
\end{itemize}

\subsection{RQ2: Retrieval of Functionality and Authorship}

RQ2 uses Codeforces and GCJ datasets to evaluate the same two representational axes through nearest-neighbor retrieval on frozen embeddings. Gallery items are ranked by cosine similarity, and retrieval performance is measured using Hit@10. Formally,
for query \(q\), let \(G_q^{(10)}\) be the first ten retrieved
gallery items and \(R_q\) be the relevant set for that query:
\begin{equation}
    \mathrm{Hit@10}(q)=\mathbf{1}\{G_q^{(10)}\cap R_q\neq\emptyset\}.
\end{equation}
We average this value over queries, and use language-macro averages for author retrieval. 

\phead{Cross-Language Functional Retrieval.}
Each implementation is used as a query to retrieve functionally equivalent implementations in the other two programming languages.
\begin{itemize}[leftmargin=*, topsep=0pt]
    \item \textbf{Query and gallery}: Each query is an implementation in one language. The gallery contains retained implementations in the other languages.
    \item \textbf{Relevance}: The relevant items are retained implementations from the same Codeforces task as the query, sharing its problem-level algorithm tags.
    \item \textbf{Ranking}: Embeddings are normalized and ranked by cosine similarity. Performance is summarized with Hit@10.
\end{itemize}

\phead{Cross-Task Author Retrieval.}
Each implementation is used to retrieve other submissions written by the same author.
\begin{itemize}[leftmargin=*, topsep=0pt]
    \item \textbf{Query and gallery}: Each query is one implementation. The gallery contains retained implementations in the query language.
    \item \textbf{Relevance}: The relevant items are retained implementations by the query author.
    \item \textbf{Ranking}: Embeddings are normalized and ranked by cosine similarity. Performance is summarized with language-macro Hit@10.
\end{itemize}

\subsection{RQ3: Controlled Analysis of Implementation Changes}

RQ3 evaluates representational invariance using the controlled implementation transformations described in Section~\ref{subsec:dataset}. For each original program $x$ from HumanEval-X and its transformed counterpart $x'$, we encode both using the same frozen model $f(\cdot)$ and compute
%
\[
    \mathrm{shift}(x,x')=\bigl(1-\cos(f(x),f(x'))\bigr)\times 100.
\]

Larger shifts indicate greater sensitivity to the applied implementation transformation. 
For each model, we summarize embedding shifts by language, transformation family, and edit magnitude.


\subsection{Experimental Settings}

\phead{Model Selection.}
We evaluate 14 embedding models: four code-focused models and ten general-purpose models. Table~\ref{tab:model-pool} reports their official names, disclosed parameter counts, and release years.
The pool spans release years from 2022 to 2025 and includes both compact encoder-scale models and billion-parameter embedders.
Disclosed parameter counts range from roughly 0.1\,B to 8\,B (Table~\ref{tab:model-pool}), while some closed-source models did not disclose parameter counts.

\begin{table}[t]
    \caption{Unified model pool. Models are grouped by official positioning and sorted by release year within each group. Parameter counts follow official documentation when available; a dash indicates no public parameter count.}
    \label{tab:model-pool}
    \small
    \centering
    \begin{tabular}{p{5.8cm}cc}
        \toprule
        \textbf{Official model name (alias)}                                                                            & \textbf{Params.} & \textbf{Year} \\
        \midrule
        \rowcolor{grouphead}\multicolumn{3}{l}{\textbf{Code embeddings}}                                                                                   \\
        UniXCoder (\textit{UniXcoder})~\cite{unixcoder}                                                                 & 125M             & 2022          \\
        voyage-code-3 (\textit{Voyage-Code-3})~\cite{voyageai2024voyagecode3}                                           & --               & 2024          \\
        CodeR (\textit{CodeR})~\cite{li2025towards}                                                                     & 1.5B             & 2025          \\
        jina-code-embeddings-1.5b (\textit{Jina-Code-1.5B})~\cite{kryvosheieva2025efficientcodeembeddingscode}          & 1.5B             & 2025          \\
        \rowcolor{grouphead}\multicolumn{3}{l}{\textbf{General-purpose embeddings}}                                                                        \\
        text-embedding-ada-002 (\textit{Ada-002})~\cite{openai2022textembeddingada002}                                  & --               & 2022          \\
        E5-Base (\textit{E5-Base})~\cite{wang2022text}                                                                  & 0.1B             & 2022          \\
        BGE-M3 (\textit{BGE-M3})~\cite{bge-m3}                                                                          & 560M             & 2024          \\
        doubao-embedding-large (\textit{Doubao-Large})~\cite{bytedance2024doubaoembeddinglarge}                         & --               & 2024          \\
        text-embedding-3-large (\textit{TE3-Large})~\cite{openai2024textembedding3}                                     & --               & 2024          \\
        text-embedding-3-small (\textit{TE3-Small})~\cite{openai2024textembedding3}                                     & --               & 2024          \\
        Llama-Embed-Nemotron-8B (\textit{Llama-Embed-8B})~\cite{babakhin2025llamaembednemotron8buniversaltextembedding} & 8B               & 2025          \\
        Qwen3-Embedding-0.6B (\textit{Qwen3-0.6B})~\cite{zhang2025qwen3embeddingadvancingtext}                          & 0.6B             & 2025          \\
        Qwen3-Embedding-4B (\textit{Qwen3-4B})~\cite{zhang2025qwen3embeddingadvancingtext}                              & 4B               & 2025          \\
        Qwen3-Embedding-8B (\textit{Qwen3-8B})~\cite{zhang2025qwen3embeddingadvancingtext}                              & 8B               & 2025          \\
        \bottomrule
    \end{tabular}
    \vspace{-3mm}
\end{table}

\phead{Encoding and Probe Training.}
Each benchmark instance is encoded once using the documented model interface, with backbone parameters frozen. For each supervised task and language, all models share the same input formulation, split, single-hidden-layer head\footnote{Our goal is to evaluate code embeddings rather than optimize downstream performance. We therefore keep the probing classifier deliberately simple to minimize confounding variables.}, 100-epoch training budget, and validation-based checkpoint rule; the embedding is the only changing factor. Each supervised run is repeated five times with independent seeds, and the reported value is the average. Optimizer settings, batch sizes, and seed lists are fixed across models. RQ2 and RQ3 fit no classifier, using cosine-similarity retrieval or scaled cosine-distance scores on frozen embeddings.

\phead{Leakage Control.}
The benchmark uses problem-level splits, per-author de-duplication, and same-task exclusion where applicable. These controls reduce leakage within the constructed evaluations. Pre-training exposure to public corpora cannot be ruled out, so we disclose source identities and model release years (Table~\ref{tab:model-pool}) rather than claim contamination-free evaluation.

\section{Results}


\subsection{RQ1: Probing Shows an Abstraction--Fidelity Split}

\begin{table}[t]
    \caption{Algorithm-tag prediction sample-F1 by language. Rows are sorted by the three-language average.}
    \label{tab:rq1-alg}
    \centering
    \small
    \setlength{\tabcolsep}{7pt}
    \renewcommand{\arraystretch}{1.06}
    \begin{tabular}{@{}lccc@{}}
        \toprule
        \textbf{Model} & \textbf{C++} & \textbf{Java} & \textbf{Python} \\
        \midrule
        Doubao-Large   & \textbf{0.542}        & 0.505         & \textbf{0.518}           \\
        CodeR          & 0.518        & \textbf{0.532}         & 0.512           \\
        Voyage-Code-3  & 0.514        & 0.530         & 0.515           \\
        Qwen3-4B       & 0.490        & 0.512         & 0.488           \\
        Qwen3-8B       & 0.487        & 0.504         & 0.498           \\
        Jina-Code-1.5B & 0.484        & 0.497         & 0.495           \\
        Qwen3-0.6B     & 0.453        & 0.455         & 0.455           \\
        Llama-Embed-8B & 0.445        & 0.446         & 0.462           \\
        Ada-002        & 0.471        & 0.414         & 0.441           \\
        TE3-Large      & 0.429        & 0.413         & 0.448           \\
        TE3-Small      & 0.404        & 0.385         & 0.405           \\
        UniXcoder      & 0.385        & 0.384         & 0.405           \\
        E5-Base        & 0.371        & 0.368         & 0.385           \\
        BGE-M3         & 0.356        & 0.348         & 0.395           \\
        \bottomrule
    \end{tabular}
    \vspace{-4mm}
\end{table}

RQ1 reveals a fundamental trade-off in frozen code embeddings: while both high-level functional intention and low-level implementation fidelity are recoverable, they are optimally supported by entirely different model profiles. At a high level, the results demonstrate that the embedding capturing algorithmic properties tend to suppress stylistic authorship signals, and vice versa. Across all 14 evaluated models, the Pearson correlation between the F1 scores (averaged across all target languages) for \emph{algorithm tag prediction} and \emph{author attribution} is negative ($r=-0.473$). This inverse relationship persists robustly when calculated within each individual language, yielding correlations of $r=-0.392$ for C++, $r=-0.583$ for Java, and $r=-0.398$ for Python. Additional Pearson tests, paired-bootstrap BCa confidence intervals, and Spearman rank correlations are provided in our anonymous repository~\cite{codeembedbench2026}. Rather than suggesting that one signal is completely absent when the other is present, these consistent negative correlations provide strong model-ranking evidence that optimization for one property inherently compromises the other. 



A closer examination of the model rankings further illustrates this trade-off. As reported in Table~\ref{tab:rq1-alg}, certain encoders make functional concepts particularly accessible to a shallow probe, with Doubao-Large achieving the highest \emph{Algorithm-Tag} Prediction scores for C++ and Python, while CodeR leads for Java. The rankings change markedly for \emph{implementation fidelity}. As shown in Table~\ref{tab:rq1-auth}, Llama-Embed-8B achieves the best performance on C++, whereas TE3-Large leads on both Java and Python. Performance on this task also spans a much wider range, from near-perfect classification to a macro-F1 of only 0.417. Notably, these leaders are not uniformly code-specialized models: general-purpose embeddings such as Doubao-Large, TE3-Large, and Llama-Embed-8B match or exceed code-focused models on these probes. 


\begin{table}[t]
    \caption{Author-attribution macro-F1 by language. Rows are sorted by the three-language average. }
    \label{tab:rq1-auth}
    \centering
    \small
    \setlength{\tabcolsep}{7pt}
    \renewcommand{\arraystretch}{1.06}
    \begin{tabular}{@{}lccc@{}}
        \toprule
        \textbf{Model} & \textbf{C++} & \textbf{Java} & \textbf{Python} \\
        \midrule
        TE3-Large      & 0.988        & \textbf{0.977}         & \textbf{0.978}           \\
        Llama-Embed-8B & \textbf{0.990}        & 0.969         & 0.945           \\
        Ada-002        & 0.988        & 0.935         & 0.940           \\
        TE3-Small      & 0.968        & 0.962         & 0.926           \\
        BGE-M3         & 0.917        & 0.954         & 0.924           \\
        Doubao-Large   & 0.923        & 0.924         & 0.921           \\
        E5-Base        & 0.920        & 0.909         & 0.869           \\
        UniXcoder      & 0.938        & 0.862         & 0.821           \\
        CodeR          & 0.871        & 0.799         & 0.849           \\
        Voyage-Code-3  & 0.800        & 0.817         & 0.826           \\
        Qwen3-0.6B     & 0.668        & 0.599         & 0.595           \\
        Qwen3-4B       & 0.587        & 0.459         & 0.417           \\
        Qwen3-8B       & 0.554        & 0.425         & 0.465           \\
        Jina-Code-1.5B & 0.521        & 0.485         & 0.430           \\
        \bottomrule
    \end{tabular}
    \vspace{-6mm}
\end{table}



The divergence becomes even clearer when considering the top-performing models. Across all three programming languages, the top-five models for the two probing tasks have no overlap. Because the two tasks use different label spaces and F1 averaging schemes, this observation is meaningful not in terms of absolute performance but in terms of relative model rankings. An engineer selecting an encoder for functional tag prediction would therefore overlook the strongest models for author attribution, and vice versa. Taken together, these results show that functional abstraction and implementation fidelity form competing representational axes in frozen code embeddings, implying that embedding model's selection should be guided by the software property most relevant to the downstream application.


\begin{finding}
    \textbf{Probing reveals a clear abstraction–fidelity split.} Both signals are recoverable, but their optimal model profiles are disjoint across all three languages. Moreover, code-specialized embeddings do not uniformly dominate general-purpose embeddings, especially on implementation-fidelity probes. Encoder selection should therefore be guided by the target software property rather than by model positioning alone.
\end{finding}

\begin{figure*}[!t]
    \centering
    \vspace{-1mm}
    \includegraphics[width=0.83\textwidth]{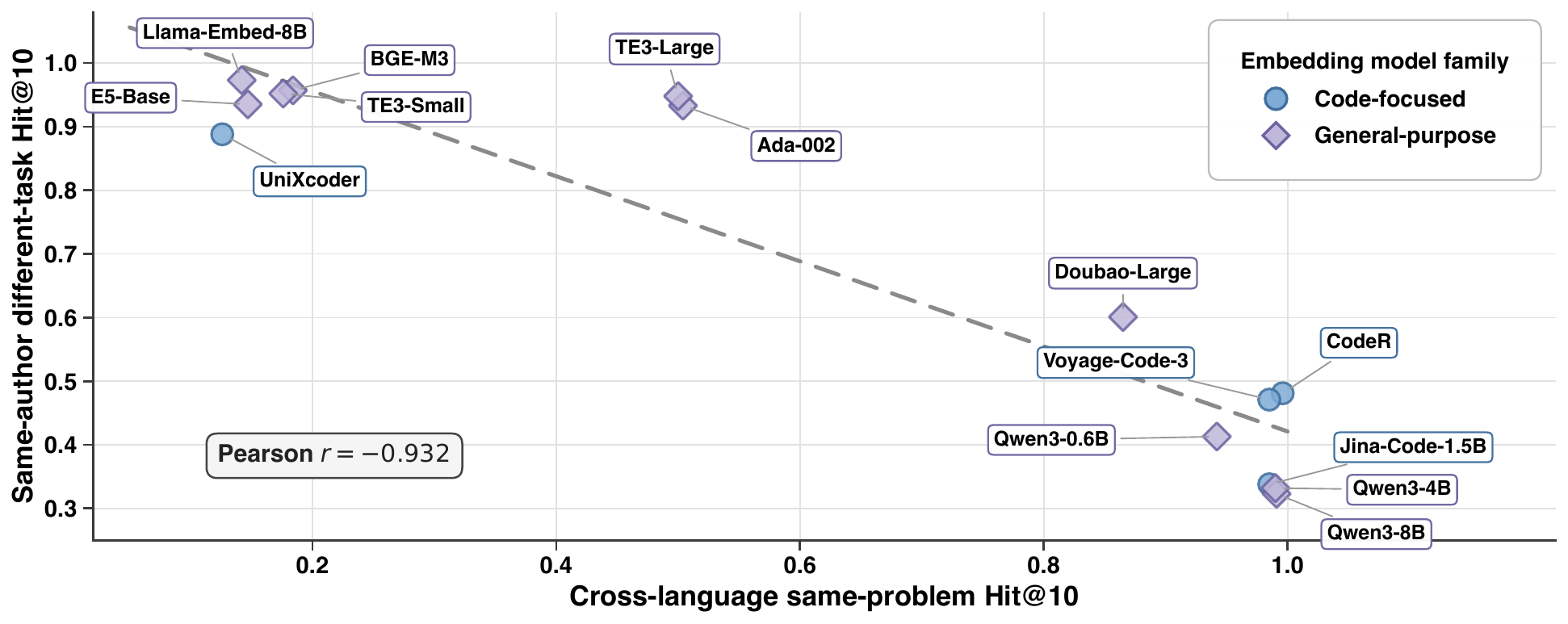}
    \vspace{-1mm}
    \caption{Same-problem versus same-author Hit@10 across the 14 models. Circles denote code-focused models, squares denote general-purpose models, the line is a linear fit.}
    \label{fig:rq2-scatter}
    \vspace{-3mm}
\end{figure*}

We also examine whether embedding models exhibit systematic preferences across programming languages. Overall, the relative model rankings are highly consistent across C++, Java, and Python for both probing tasks, suggesting that the abstraction–fidelity trade-off is largely language independent. For example, Doubao-Large, CodeR, and Voyage-Code-3 consistently rank among the strongest models for Algorithm-Tag Prediction, whereas TE3-Large and Llama-Embed-8B remain the strongest performers for Author Attribution across all three languages. The absolute performance differences between languages are also modest for most models, indicating that representational preferences are driven primarily by the embedding model rather than the programming language. 

\subsection{RQ2: Retrieval Shows a Stronger Abstraction--Fidelity Split}


\begin{table}[!t]
    \caption{Hit@10 for cross-language retrieval. Functional retrieval ranks implementations in other languages; author retrieval ranks same-author implementations.}
    \label{tab:rq2-retrieval-all}
    \centering
    \small
    \setlength{\tabcolsep}{3.6pt}
    \renewcommand{\arraystretch}{1.08}
    \begin{tabular}{@{}lcc@{}}
        \toprule
        \textbf{Model} & \textbf{Functional} & \textbf{Author} \\
        \midrule
        CodeR          & 0.996               & 0.481           \\
        Qwen3-8B       & 0.991               & 0.323           \\
        Qwen3-4B       & 0.990               & 0.332           \\
        Voyage-Code-3  & 0.985               & 0.471           \\
        Jina-Code-1.5B & 0.985               & 0.338           \\
        Qwen3-0.6B     & 0.942               & 0.413           \\
        Doubao-Large   & 0.865               & 0.601           \\
        Ada-002        & 0.504               & 0.933           \\
        TE3-Large      & 0.500               & 0.948           \\
        BGE-M3         & 0.184               & 0.957           \\
        TE3-Small      & 0.176               & 0.952           \\
        E5-Base        & 0.147               & 0.935           \\
        Llama-Embed-8B & 0.142               & 0.973           \\
        UniXcoder      & 0.126               & 0.888           \\
        \bottomrule
    \end{tabular}
\end{table}

While RQ1 demonstrates that both computational-intent and implementation-realization evidence can be recovered through lightweight probes, RQ2 moves to a more practical setting by evaluating embeddings directly in retrieval, without relying on a trained prediction head. If a single embedding space supported both representational axes equally well, functionally related programs and implementation-aligned programs by the same author would both appear among a query’s nearest neighbors. Instead, direct retrieval reveals a pronounced separation between these two objectives.

This separation is reflected consistently across models. Across the 14 evaluated models, Cross-Language Functional Retrieval and Cross-Task Author Retrieval are strongly negatively correlated ($r=-0.932$; Table~\ref{tab:rq2-retrieval-all}, Figure~\ref{fig:rq2-scatter}). The evaluated models fall into two specialized groups. Functional specialists, including CodeR, Qwen3-8B, Qwen3-4B, Voyage-Code-3, and Jina-Code-1.5B, excel at semantic retrieval (Hit@10 $\ge 0.985$) but perform substantially worse on author retrieval (Hit@10: 0.323–0.481). In contrast, author-oriented models such as BGE-M3, TE3-Small, E5-Base, Llama-Embed-8B, and UniXcoder achieve high author retrieval accuracy (Hit@10: 0.888–0.973) while exhibiting poor cross-language functional retrieval (Hit@10 $\le 0.184$). This geometric split can directly impact downstream utility. Selecting a model for cross-language functional search pushes authorial signals out of the local neighborhood, while optimizing for authorship tracking largely removes functionally related implementations across different programming languages.

\begin{finding}
    \textbf{Retrieval makes the abstraction--fidelity split clear.}
    Evaluation results expose a clearer trade-off between the two axes. Across the 14 evaluated models, functional and author retrieval performance is strongly negatively correlated ($r=-0.932$). Consequently, we argue that embedding selection should therefore be guided by the task property rather than a single overall ranking.
\end{finding}

\begin{table*}[!t]
    \caption{Scaled cosine distance from each original program to its test-passing transformed version, averaged over languages. \emph{low} ($0\le r\le0.33$), \emph{med.} ($0.33<r\le0.66$), and \emph{high} ($r>0.66$) denote transformation strength.}
    \label{tab:rq3-rewrite-detail}
    \centering
    \small
    \renewcommand{\arraystretch}{1.05}
    \setlength{\tabcolsep}{3.3pt}
    \begin{tabular}{lrrrrrrrrrrrrrrrrr}
        \toprule
                       & \multicolumn{3}{c}{Data/Decl.} & \multicolumn{3}{c}{Control} & \multicolumn{3}{c}{Function} & \multicolumn{3}{c}{Dead/Bogus} & \multicolumn{3}{c}{Format/Rename}                                                                                        \\
        \cmidrule(lr){2-4}\cmidrule(lr){5-7}\cmidrule(lr){8-10}\cmidrule(lr){11-13}\cmidrule(l){14-16}
        Model          & \multicolumn{1}{c}{\emph{low}}                              & \multicolumn{1}{c}{\emph{med.}}                           & \multicolumn{1}{c}{\emph{high}}                            & \multicolumn{1}{c}{\emph{low}}                              & \multicolumn{1}{c}{\emph{med.}}                                 & \multicolumn{1}{c}{\emph{high}}      & \multicolumn{1}{c}{\emph{low}}     & \multicolumn{1}{c}{\emph{med.}}      & \multicolumn{1}{c}{\emph{high}}      & \multicolumn{1}{c}{\emph{low}}     & \multicolumn{1}{c}{\emph{med.}}      & \multicolumn{1}{c}{\emph{high}}      & \multicolumn{1}{c}{\emph{low}}     & \multicolumn{1}{c}{\emph{med.}}      & \multicolumn{1}{c}{\emph{high}}      \\
        \midrule
        CodeR          & 1.717                          & 4.601                       & 10.209                       & 1.460                          & 3.344                             & 7.282  & 4.227 & 7.838  & 15.403 & 4.123 & 10.032 & 22.136 & 3.262 & 5.926  & 12.204 \\
        Qwen3-8B       & 1.085                          & 2.481                       & 4.392                        & 1.013                          & 2.033                             & 3.656  & 1.688 & 3.355  & 6.434  & 2.373 & 5.263  & 8.684  & 2.805 & 4.268  & 7.620  \\
        Qwen3-4B       & 1.090                          & 2.392                       & 4.168                        & 0.995                          & 2.039                             & 3.479  & 1.725 & 3.465  & 6.416  & 2.086 & 4.919  & 10.526 & 3.134 & 4.589  & 7.407  \\
        Voyage-Code-3  & 0.605                          & 1.704                       & 3.794                        & 0.558                          & 1.390                             & 2.853  & 0.959 & 2.321  & 5.370  & 1.733 & 3.881  & 8.935  & 1.572 & 2.413  & 5.080  \\
        Jina-Code-1.5B & 1.706                          & 5.181                       & 10.053                       & 1.269                          & 3.121                             & 6.888  & 2.617 & 6.426  & 14.370 & 3.942 & 11.361 & 27.248 & 2.670 & 4.063  & 9.306  \\
        \midrule
        Qwen3-0.6B     & 1.127                          & 2.859                       & 4.887                        & 0.813                          & 1.712                             & 2.961  & 1.535 & 3.079  & 5.846  & 1.804 & 4.656  & 9.859  & 2.421 & 3.496  & 5.951  \\
        Doubao-Large   & 0.807                          & 1.654                       & 3.191                        & 0.791                          & 1.653                             & 2.804  & 1.200 & 2.275  & 4.693  & 1.419 & 2.948  & 5.679  & 1.551 & 2.202  & 4.202  \\
        Ada-002        & 0.600                          & 1.381                       & 2.978                        & 0.469                          & 0.947                             & 1.916  & 1.273 & 2.830  & 5.863  & 1.063 & 2.489  & 6.370  & 1.863 & 2.475  & 4.752  \\
        TE3-Large      & 2.522                          & 5.857                       & 11.413                       & 2.189                          & 4.225                             & 7.622  & 4.652 & 9.403  & 17.683 & 4.716 & 8.629  & 17.870 & 5.238 & 7.616  & 13.785 \\
        \midrule
        BGE-M3         & 1.481                          & 4.527                       & 11.204                       & 1.225                          & 2.826                             & 6.424  & 2.754 & 7.069  & 16.882 & 4.711 & 8.768  & 19.530 & 3.648 & 7.459  & 14.128 \\
        TE3-Small      & 1.292                          & 3.189                       & 7.482                        & 1.008                          & 2.062                             & 4.508  & 4.327 & 10.304 & 19.787 & 2.544 & 5.971  & 16.709 & 4.391 & 5.865  & 11.091 \\
        E5-Base        & 0.630                          & 1.883                       & 5.571                        & 0.580                          & 1.298                             & 3.088  & 1.456 & 3.516  & 8.377  & 1.911 & 4.049  & 9.607  & 1.357 & 2.620  & 6.070  \\
        Llama-Embed-8B & 2.891                          & 6.993                       & 16.877                       & 2.596                          & 5.277                             & 11.261 & 4.358 & 10.347 & 24.398 & 5.069 & 11.359 & 30.738 & 5.070 & 8.175  & 18.494 \\
        UniXcoder      & 3.854                          & 9.130                       & 24.349                       & 3.528                          & 6.838                             & 16.678 & 6.880 & 14.428 & 37.139 & 9.484 & 20.674 & 45.312 & 6.627 & 11.692 & 28.849 \\
        \bottomrule
    \end{tabular}
    \vspace{-3mm}
\end{table*}

Interestingly, we also observe that scaling within the same embedding family can shift representational preferences. As the Qwen3 embedding model scales from 0.6B to 4B and 8B parameters, functional retrieval steadily improves (Hit@10: 0.942 → 0.991), while author retrieval declines correspondingly (Hit@10: 0.413 → 0.323). This trend mirrors the probing results from RQ1, where larger Qwen3 variants achieved higher Algorithm-Tag Prediction performance at the expense of Author Attribution. Taken together, these findings suggest that increasing model scale strengthens representations of computational intent while gradually reducing the prominence of implementation-level information in the embedding space, particularly in the nearest-neighbor geometry.




\subsection{RQ3: Controlled Rewrites Reveal Selective Invariance}


In RQ3, we treat each functionality-preserving mutation as a counterfactual intervention: the program's tested behavior is held constant while specific aspects of its implementation are modified. Small embedding displacements indicate that the altered implementation detail has been abstracted away, whereas larger displacements indicate that it remains geometrically salient.
%
%
As expected from the edit-strength levels defined in Section~\ref{subsec:dataset}, geometric displacement increases monotonically from low- to medium- to high-edit-magnitude transformations across every model and transformation family (Table~\ref{tab:rq3-rewrite-detail}), confirming that the metric responds consistently to progressively larger implementation changes.
We therefore focus on how displacement differs across implementation transformation families and embedding models.


The results provide a geometric explanation for the abstraction–fidelity trade-off observed in RQ1 and RQ2. Models that excel at functional retrieval consistently exhibit smaller embedding shifts under all five transformation families than models that perform poorly on functional retrieval. 
The largest difference occurs for Function rewrites, where the average displacement is 5.508 for the high-functional group compared with 11.468 for the low-functional group. These results suggest that embeddings optimized for functional retrieval are more invariant to implementation-level variation.


However, this invariance is highly selective rather than uniform. Across all 14 models, Control-Flow transformations consistently produce the smallest geometric shifts. For example, under high-edit-magnitude transformations, Voyage-Code-3, Doubao-Large, and Qwen3-8B exhibit shifts of only 2.853, 2.804, and 3.656, respectively. In contrast, Dead/Bogus Code and Function transformations are consistently the most disruptive. UniXcoder, for example, reaches geometric shifts of 45.312 for Dead/Bogus Code and 37.139 for Function transformations, while Llama-Embed-8B reaches 30.738 and 24.398, respectively. These observations indicate that control-flow variations are widely abstracted away, whereas program decomposition and semantically inert code remain much more salient in the embedding space.

Surprisingly, models with comparable functional retrieval performance do not necessarily exhibit the same invariance profile. Voyage-Code-3 and Qwen3 series remain relatively insensitive across most transformation families, whereas CodeR and Jina-Code-1.5B are substantially more sensitive to Function and Dead/Bogus transformations despite achieving similarly strong retrieval performance. 
This suggests that comparable functional abstraction can arise through different representational strategies rather than a single notion of invariance.


\begin{finding}
    \textbf{Functional abstraction is achieved through selective implementation invariance rather than indiscriminate robustness.} Embeddings optimized for functional retrieval consistently suppress implementation-level variation, but different models abstract away different types of implementation changes, indicating multiple representational strategies for achieving similar retrieval behavior.
\end{finding}

\section{Discussion and Future Work}


Our findings shed light on several future directions for code embedding models in software engineering.

\phead{Rethinking the Evaluation of Code Embedding Models.}
Current evaluations of code embedding models are typically organized around one or a small number of downstream tasks, implicitly treating strong benchmark performance as evidence of a generally useful representation. Our results suggest that this assumption should be revisited. Across probing (RQ1), retrieval (RQ2), and controlled transformation analysis (RQ3), we consistently observe that embedding models preserve different software properties to different degrees. Models that perform strongly on functional abstraction often rank poorly on implementation-sensitive tasks, while models with similar retrieval performance can nevertheless exhibit substantially different responses to implementation changes. These findings indicate that downstream performance alone provides only a partial view of an embedding’s representational behavior.

Building on these observations, future evaluation frameworks may characterize a broader range of software relationships than those considered in this study. While we examine functional abstraction, implementation fidelity, and implementation-level invariance, embedding models may also encode behavioral, structural, evolutionary, and project-specific relationships. Future studies could therefore investigate representations of API usage, architectural roles, dependency structures, security properties, and software evolution through commits, refactorings, and code reviews. More importantly, our results suggest that embedding evaluation should become increasingly diagnostic rather than purely comparative. Instead of reporting only which model performs best, future studies may also investigate why embedding models succeed or fail by identifying which software properties are preserved, which are abstracted away, and how these representations evolve under realistic software development activities. Such evaluations would provide a richer understanding of embedding behavior while offering more actionable guidance for both model developers and practitioners.

\phead{Representation-Aware Deployment of Code Embeddings.}
Our study also has practical implications for the deployment of embedding models in software engineering systems. A common practice is to adopt a single embedding model as a general-purpose backbone for diverse downstream applications, including code search, repository retrieval, and software analytics. However, our results suggest that embedding models should not be regarded as interchangeable. Across RQ1 and RQ2, models that excel at functional abstraction consistently differ from those that preserve implementation-level information, while RQ3 further shows that even models with comparable retrieval performance may encode substantially different notions of implementation similarity.

Rather than selecting embedding models solely according to overall benchmark performance, practitioners should instead consider the software properties required by the target application. Models that strongly abstract implementation variation are naturally suited to semantic retrieval and search-oriented applications, whereas models that retain implementation-level information are more appropriate for provenance analysis, authorship attribution, and other style-sensitive tasks. More broadly, our perturbation analysis suggests that robustness to implementation variation should become an additional consideration during model selection, particularly for applications expected to operate over evolving codebases. From a sustainability perspective, representation-aware deployment also enables more effective reuse of existing embedding models. Rather than repeatedly training or deploying multiple task-specific embeddings, practitioners can select representations whose behavioral characteristics already align with application requirements. This reduces unnecessary model development, reduces computational and maintenance overhead, and ultimately contributes to more sustainable AI software.

\section{Threats to Validity}

\phead{Construct Validity.} A primary threat to construct validity is whether our evaluation tasks adequately operationalize the broader concepts of functional abstraction and implementation fidelity.  To mitigate this threat, we evaluate complementary properties through probing, retrieval, and controlled transformation analyses across multiple datasets and languages, providing converging evidence from different perspectives.

\phead{Internal Validity.} One potential threat to internal validity is that an expressive downstream classifier could overcompensate for poor embeddings, obscuring differences in representation quality. To mitigate this, all models share identical probe architectures  (a single-hidden-layer MLP), optimization settings, data splits, and training budgets so that the embedding representation remains the primary experimental variable. We also repeated each training run five times with different random seeds and report the average performance. In addition, controlled transformations are generated with an LLM and therefore may not fully represent edits produced during real software development. We mitigate this by requiring all transformed programs to pass the original test suite, validating transformation correctness using an independent LLM, and performing manual inspection on a sampled subset.

\phead{External Validity.} One threat to external validity is that our experiment tasks are constructed from competitive programming and code generation datasets, which may not fully reflect the characteristics of industrial software projects. We partially mitigate this threat by using multiple datasets that capture complementary software engineering scenarios, although future studies should validate our findings on large-scale industrial repositories. Another threat is that our evaluation is limited to function-level code written in C++, Java, and Python. Future work may consider investigating repository-level representations and additional programming languages. Finally, although we evaluated a highly diverse pool of 14 SOTA models, ranging from compact encoders to billion-parameter general-purpose embeddings, our findings may not generalize to future embedding models. Future work may extend our experiments to newer embedding architectures or models trained with substantially different objectives.


\section{Conclusion}

In this paper, we moved beyond monolithic downstream benchmarks to systematically deconstruct the latent spaces of 14 SOTA code embedding models. By evaluating these representations through the distinct lenses of supervised probing, geometric retrieval, and invariant perturbation, we exposed an inherent \emph{abstraction--fidelity tension}. Our findings demonstrate that semantic generalization and implementation-level sensitivity operate as competing design trade-offs. Furthermore, domain-specific pretraining is not a panacea, as code-specialized models do not uniformly outperform general-purpose alternatives. These insights demonstrate that a single, universal code representation remains out of reach. For practitioners, model selection should be treated as an explicit alignment problem between a task's unique requirements and an embedding's intrinsic biases. For the research community, our results highlight the limitations of aggregate, single-score leaderboards. Rather than chasing a hypothetical all-purpose vector space, future work can focus on evaluation frameworks and model architectures that explicitly characterize, partition, or dynamically adapt to these distinct code properties.



\bibliographystyle{IEEEtran}
\bibliography{ref}

\begin{thebibliography}{10}
\providecommand{\url}[1]{#1}
\csname url@samestyle\endcsname
\providecommand{\newblock}{\relax}
\providecommand{\bibinfo}[2]{#2}
\providecommand{\BIBentrySTDinterwordspacing}{\spaceskip=0pt\relax}
\providecommand{\BIBentryALTinterwordstretchfactor}{4}
\providecommand{\BIBentryALTinterwordspacing}{\spaceskip=\fontdimen2\font plus
\BIBentryALTinterwordstretchfactor\fontdimen3\font minus \fontdimen4\font\relax}
\providecommand{\BIBforeignlanguage}[2]{{%
\expandafter\ifx\csname l@#1\endcsname\relax
\typeout{** WARNING: IEEEtran.bst: No hyphenation pattern has been}%
\typeout{** loaded for the language `#1'. Using the pattern for}%
\typeout{** the default language instead.}%
\else
\language=\csname l@#1\endcsname
\fi
#2}}
\providecommand{\BIBdecl}{\relax}
\BIBdecl

\bibitem{gu2018deep}
X.~Gu, H.~Zhang, and S.~Kim, ``Deep code search,'' in \emph{Proceedings of the 40th international conference on software engineering}, 2018, pp. 933--944.

\bibitem{zhou2019devign}
Y.~Zhou, S.~Liu, J.~Siow, X.~Du, and Y.~Liu, ``Devign: Effective vulnerability identification by learning comprehensive program semantics via graph neural networks,'' \emph{Advances in neural information processing systems}, vol.~32, 2019.

\bibitem{husain2019codesearchnet}
H.~Husain, H.-H. Wu, T.~Gazit, M.~Allamanis, and M.~Brockschmidt, ``Codesearchnet challenge: Evaluating the state of semantic code search,'' \emph{arXiv preprint arXiv:1909.09436}, 2019.

\bibitem{ding2022can}
Z.~Ding, H.~Li, W.~Shang, and T.-H.~P. Chen, ``Can pre-trained code embeddings improve model performance? revisiting the use of code embeddings in software engineering tasks,'' \emph{Empirical Software Engineering}, vol.~27, no.~3, p.~63, 2022.

\bibitem{xian2024transformcode}
Z.~Xian, R.~Huang, D.~Towey, C.~Fang, and Z.~Chen, ``Transformcode: A contrastive learning framework for code embedding via subtree transformation,'' \emph{IEEE Transactions on Software Engineering}, vol.~50, no.~6, pp. 1600--1619, 2024.

\bibitem{wang2026towards}
\BIBentryALTinterwordspacing
Y.~Wang, K.~Duan, D.~Zheng, E.~Shi, F.~Zhang, Y.~Wang, J.~Chen, X.~Liu, Y.~Ma, H.~Zhang, Q.~Wang, and Z.~Zheng, ``Towards an understanding of context utilization in code intelligence,'' vol.~58, no.~11, Apr. 2026. [Online]. Available: \url{https://doi.org/10.1145/3797261}
\BIBentrySTDinterwordspacing

\bibitem{vuldeepecker}
\BIBentryALTinterwordspacing
Z.~Li, D.~Zou, S.~Xu, X.~Ou, H.~Jin, S.~Wang, Z.~Deng, and Y.~Zhong, ``Vuldeepecker: A deep learning-based system for vulnerability detection.'' in \emph{NDSS}.\hskip 1em plus 0.5em minus 0.4em\relax The Internet Society, 2018. [Online]. Available: \url{http://dblp.uni-trier.de/db/conf/ndss/ndss2018.html#LiZXO0WDZ18}
\BIBentrySTDinterwordspacing

\bibitem{wen2025repository}
X.-C. Wen, Z.~Lin, C.~Gao, H.~Zhang, Y.~Wang, and Q.~Liao, ``Repository-level graph representation learning for enhanced security patch detection,'' in \emph{2025 IEEE/ACM 47th International Conference on Software Engineering (ICSE)}.\hskip 1em plus 0.5em minus 0.4em\relax IEEE, 2025, pp. 1--13.

\bibitem{zhang2019novel}
J.~Zhang, X.~Wang, H.~Zhang, H.~Sun, K.~Wang, and X.~Liu, ``A novel neural source code representation based on abstract syntax tree,'' in \emph{Proceedings of the 41st International Conference on Software Engineering}.\hskip 1em plus 0.5em minus 0.4em\relax IEEE Press, 2019, pp. 783--794.

\bibitem{hoang2020cc2vec}
T.~Hoang, H.~J. Kang, D.~Lo, and J.~Lawall, ``Cc2vec: Distributed representations of code changes,'' in \emph{Proceedings of the ACM/IEEE 42nd international conference on software engineering}, 2020, pp. 518--529.

\bibitem{codebert}
\BIBentryALTinterwordspacing
Z.~Feng, D.~Guo, D.~Tang, N.~Duan, X.~Feng, M.~Gong, L.~Shou, B.~Qin, T.~Liu, D.~Jiang, and M.~Zhou, ``{C}ode{BERT}: A pre-trained model for programming and natural languages,'' in \emph{Findings of the Association for Computational Linguistics: EMNLP 2020}, T.~Cohn, Y.~He, and Y.~Liu, Eds.\hskip 1em plus 0.5em minus 0.4em\relax Online: Association for Computational Linguistics, Nov. 2020, pp. 1536--1547. [Online]. Available: \url{https://aclanthology.org/2020.findings-emnlp.139/}
\BIBentrySTDinterwordspacing

\bibitem{yan2024enhancing}
Y.~Yan, N.~Cooper, K.~Moran, G.~Bavota, D.~Poshyvanyk, and S.~Rich, ``Enhancing code understanding for impact analysis by combining transformers and program dependence graphs,'' \emph{Proceedings of the ACM on Software Engineering}, vol.~1, no. FSE, pp. 972--995, 2024.

\bibitem{wang-etal-2025-coderag}
\BIBentryALTinterwordspacing
Z.~Z. Wang, A.~Asai, X.~V. Yu, F.~F. Xu, Y.~Xie, G.~Neubig, and D.~Fried, ``{C}ode{RAG}-bench: Can retrieval augment code generation?'' in \emph{Findings of the Association for Computational Linguistics: NAACL 2025}, L.~Chiruzzo, A.~Ritter, and L.~Wang, Eds.\hskip 1em plus 0.5em minus 0.4em\relax Albuquerque, New Mexico: Association for Computational Linguistics, Apr. 2025, pp. 3199--3214. [Online]. Available: \url{https://aclanthology.org/2025.findings-naacl.176/}
\BIBentrySTDinterwordspacing

\bibitem{tao2025retrieval}
Y.~Tao, Y.~Li, Y.~Qin, and Y.~Liu, ``Retrieval-augmented code generation: A survey with focus on repository-level approaches,'' \emph{arXiv preprint arXiv:2510.04905}, 2025.

\bibitem{zeng2023degraphcs}
C.~Zeng, Y.~Yu, S.~Li, X.~Xia, Z.~Wang, M.~Geng, L.~Bai, W.~Dong, and X.~Liao, ``degraphcs: Embedding variable-based flow graph for neural code search,'' \emph{ACM Transactions on Software Engineering and Methodology}, vol.~32, no.~2, pp. 1--27, 2023.

\bibitem{das2023comex}
D.~Das, N.~S. Mathews, A.~Mathai, S.~Tamilselvam, K.~Sedamaki, S.~Chimalakonda, and A.~Kumar, ``Comex: a tool for generating customized source code representations,'' in \emph{2023 38th IEEE/ACM International Conference on Automated Software Engineering (ASE)}.\hskip 1em plus 0.5em minus 0.4em\relax IEEE, 2023, pp. 2054--2057.

\bibitem{yang2023evacrc}
L.~Yang, J.~Xu, Y.~Zhang, H.~Zhang, and A.~Bacchelli, ``Evacrc: Evaluating code review comments,'' in \emph{Proceedings of the 31st ACM Joint European Software Engineering Conference and Symposium on the Foundations of Software Engineering}, 2023, pp. 275--287.

\bibitem{zhao2026understanding}
S.~Zhao, Y.~Shao, Y.~Huang, J.~Song, Z.~Wang, C.~Wan, and L.~Ma, ``Understanding the fundamental design decisions of retrieval-augmented generation systems,'' \emph{ACM Transactions on Software Engineering and Methodology}, 2026.

\bibitem{kang2019assessing}
H.~J. Kang, T.~F. Bissyand{\'e}, and D.~Lo, ``Assessing the generalizability of code2vec token embeddings,'' in \emph{2019 34th IEEE/ACM International Conference on Automated Software Engineering (ASE)}.\hskip 1em plus 0.5em minus 0.4em\relax IEEE, 2019, pp. 1--12.

\bibitem{huang2025look}
Y.~Huang, J.~Song, Z.~Wang, S.~Zhao, H.~Chen, F.~Juefei-Xu, and L.~Ma, ``Look before you leap: An exploratory study of uncertainty analysis for large language models,'' \emph{IEEE Transactions on Software Engineering}, vol.~51, no.~2, pp. 413--429, 2025.

\bibitem{fein2026challenges}
B.~Fein, M.~Jungwirth, G.~Fraser, and F.~Kandlinger, ``Challenges of deploying code embeddings: an industrial case study on method name generation,'' \emph{Automated Software Engineering}, vol.~33, no.~2, p.~53, 2026.

\bibitem{Ribeiro2026CodeCorrectness}
F.~Ribeiro, C.~Spiess, P.~Devanbu, and S.~Nadi, ``On {LLMs'} internal representation of code correctness,'' in \emph{Proceedings of the 48th International Conference on Software Engineering (ICSE '26)}.\hskip 1em plus 0.5em minus 0.4em\relax Rio de Janeiro, Brazil: Association for Computing Machinery, 2026.

\bibitem{neelakantan2022text}
A.~Neelakantan, T.~Xu, R.~Puri, A.~Radford, J.~M. Han, J.~Tworek, Q.~Yuan, N.~Tezak, J.~W. Kim, C.~Hallacy \emph{et~al.}, ``Text and code embeddings by contrastive pre-training,'' \emph{arXiv preprint arXiv:2201.10005}, 2022.

\bibitem{wang2022text}
L.~Wang, N.~Yang, X.~Huang, B.~Jiao, L.~Yang, D.~Jiang, R.~Majumder, and F.~Wei, ``Text embeddings by weakly-supervised contrastive pre-training,'' \emph{arXiv preprint arXiv:2212.03533}, 2022.

\bibitem{bge-m3}
J.~Chen, S.~Xiao, P.~Zhang, K.~Luo, D.~Lian, and Z.~Liu, ``Bge m3-embedding: Multi-lingual, multi-functionality, multi-granularity text embeddings through self-knowledge distillation,'' 2024.

\bibitem{muennighoff2023mteb}
N.~Muennighoff, N.~Tazi, L.~Magne, and N.~Reimers, ``Mteb: Massive text embedding benchmark,'' in \emph{Proceedings of the 17th Conference of the European Chapter of the Association for Computational Linguistics}, 2023, pp. 2014--2037.

\bibitem{liu2025codexembed}
\BIBentryALTinterwordspacing
Y.~Liu, R.~Meng, S.~Joty, silvio savarese, C.~Xiong, Y.~Zhou, and S.~Yavuz, ``Code{XE}mbed: A generalist embedding model family for multilingual and multi-task code retrieval,'' in \emph{Second Conference on Language Modeling}, 2025. [Online]. Available: \url{https://openreview.net/forum?id=z3lG70Azbg}
\BIBentrySTDinterwordspacing

\bibitem{li2025towards}
\BIBentryALTinterwordspacing
C.~Li, J.~Chen, Y.~Shao, D.~Lian, and Z.~Liu, ``Towards a generalist code embedding model based on massive data synthesis,'' in \emph{The Thirty-ninth Annual Conference on Neural Information Processing Systems Datasets and Benchmarks Track}, 2025. [Online]. Available: \url{https://openreview.net/forum?id=K3S0uNfUsY}
\BIBentrySTDinterwordspacing

\bibitem{tian2020evaluating}
H.~Tian, K.~Liu, A.~K. Kabor{\'e}, A.~Koyuncu, L.~Li, J.~Klein, and T.~F. Bissyand{\'e}, ``Evaluating representation learning of code changes for predicting patch correctness in program repair,'' in \emph{Proceedings of the 35th IEEE/ACM international conference on automated software engineering}, 2020, pp. 981--992.

\bibitem{zhao2024codingptms}
Y.~Zhao, L.~Gong, Z.~Huang, Y.~Wang, M.~Wei, and F.~Wu, ``Coding-ptms: How to find optimal code pre-trained models for code embedding in vulnerability detection?'' in \emph{Proceedings of the 39th IEEE/ACM International Conference on Automated Software Engineering}, 2024, pp. 1732--1744.

\bibitem{codeembedbench2026}
{Anonymous Authors}, ``Our website,'' \url{https://anonymous.4open.science/r/CodeEmbedBench-816922/}, 2026, anonymous repository.

\bibitem{svajlenko2014towards}
J.~Svajlenko, J.~F. Islam, I.~Keivanloo, C.~K. Roy, and M.~M. Mia, ``Towards a big data curated benchmark of inter-project code clones,'' in \emph{2014 IEEE international conference on software maintenance and evolution}.\hskip 1em plus 0.5em minus 0.4em\relax IEEE, 2014, pp. 476--480.

\bibitem{PLBART}
\BIBentryALTinterwordspacing
W.~Ahmad, S.~Chakraborty, B.~Ray, and K.-W. Chang, ``Unified pre-training for program understanding and generation,'' in \emph{Proceedings of the 2021 Conference of the North American Chapter of the Association for Computational Linguistics: Human Language Technologies}.\hskip 1em plus 0.5em minus 0.4em\relax Online: Association for Computational Linguistics, Jun. 2021, pp. 2655--2668. [Online]. Available: \url{https://www.aclweb.org/anthology/2021.naacl-main.211}
\BIBentrySTDinterwordspacing

\bibitem{word2vec}
\BIBentryALTinterwordspacing
T.~Mikolov, K.~Chen, G.~Corrado, and J.~Dean, ``Efficient estimation of word representations in vector space,'' in \emph{1st International Conference on Learning Representations, {ICLR} 2013, Scottsdale, Arizona, USA, May 2-4, 2013, Workshop Track Proceedings}, 2013. [Online]. Available: \url{http://arxiv.org/abs/1301.3781}
\BIBentrySTDinterwordspacing

\bibitem{code2vec}
\BIBentryALTinterwordspacing
U.~Alon, M.~Zilberstein, O.~Levy, and E.~Yahav, ``Code2vec: Learning distributed representations of code,'' \emph{Proc. ACM Program. Lang.}, vol.~3, no. POPL, pp. 40:1--40:29, Jan. 2019. [Online]. Available: \url{http://doi.acm.org/10.1145/3290353}
\BIBentrySTDinterwordspacing

\bibitem{code2seq}
\BIBentryALTinterwordspacing
U.~Alon, S.~Brody, O.~Levy, and E.~Yahav, ``code2seq: Generating sequences from structured representations of code,'' in \emph{International Conference on Learning Representations}, 2019. [Online]. Available: \url{https://openreview.net/forum?id=H1gKYo09tX}
\BIBentrySTDinterwordspacing

\bibitem{bui2021infercode}
N.~D. Bui, Y.~Yu, and L.~Jiang, ``Infercode: Self-supervised learning of code representations by predicting subtrees,'' in \emph{2021 IEEE/ACM 43rd International Conference on Software Engineering (ICSE)}.\hskip 1em plus 0.5em minus 0.4em\relax IEEE, 2021, pp. 1186--1197.

\bibitem{HedgeCode}
G.~Chen, X.~Xie, D.~Tang, Q.~Xin, and W.~Liu, ``Hedgecode: A multi-task hedging contrastive learning framework for code search,'' in \emph{2025 IEEE/ACM 47th International Conference on Software Engineering (ICSE)}.\hskip 1em plus 0.5em minus 0.4em\relax IEEE, 2025, pp. 1857--1868.

\bibitem{liu-etal-2023-code}
\BIBentryALTinterwordspacing
C.~Liu, S.~Lu, W.~Chen, D.~Jiang, A.~Svyatkovskiy, S.~Fu, N.~Sundaresan, and N.~Duan, ``Code execution with pre-trained language models,'' in \emph{Findings of the Association for Computational Linguistics: ACL 2023}, A.~Rogers, J.~Boyd-Graber, and N.~Okazaki, Eds.\hskip 1em plus 0.5em minus 0.4em\relax Toronto, Canada: Association for Computational Linguistics, Jul. 2023, pp. 4984--4999. [Online]. Available: \url{https://aclanthology.org/2023.findings-acl.308/}
\BIBentrySTDinterwordspacing

\bibitem{LongCoder}
D.~Guo, C.~Xu, N.~Duan, J.~Yin, and J.~McAuley, ``Longcoder: a long-range pre-trained language model for code completion,'' in \emph{Proceedings of the 40th International Conference on Machine Learning}, ser. ICML'23, 2023.

\bibitem{graphcodebert}
\BIBentryALTinterwordspacing
D.~Guo, S.~Ren, S.~Lu, Z.~Feng, D.~Tang, S.~LIU, L.~Zhou, N.~Duan, A.~Svyatkovskiy, S.~Fu, M.~Tufano, S.~K. Deng, C.~Clement, D.~Drain, N.~Sundaresan, J.~Yin, D.~Jiang, and M.~Zhou, ``Graphcode{\{}bert{\}}: Pre-training code representations with data flow,'' in \emph{International Conference on Learning Representations}, 2021. [Online]. Available: \url{https://openreview.net/forum?id=jLoC4ez43PZ}
\BIBentrySTDinterwordspacing

\bibitem{unixcoder}
\BIBentryALTinterwordspacing
D.~Guo, S.~Lu, N.~Duan, Y.~Wang, M.~Zhou, and J.~Yin, ``{U}ni{X}coder: Unified cross-modal pre-training for code representation,'' in \emph{Proceedings of the 60th Annual Meeting of the Association for Computational Linguistics (Volume 1: Long Papers)}, S.~Muresan, P.~Nakov, and A.~Villavicencio, Eds.\hskip 1em plus 0.5em minus 0.4em\relax Dublin, Ireland: Association for Computational Linguistics, May 2022, pp. 7212--7225. [Online]. Available: \url{https://aclanthology.org/2022.acl-long.499/}
\BIBentrySTDinterwordspacing

\bibitem{codet5}
\BIBentryALTinterwordspacing
Y.~Wang, W.~Wang, S.~Joty, and S.~C. Hoi, ``{C}ode{T}5: Identifier-aware unified pre-trained encoder-decoder models for code understanding and generation,'' in \emph{Proceedings of the 2021 Conference on Empirical Methods in Natural Language Processing}, M.-F. Moens, X.~Huang, L.~Specia, and S.~W.-t. Yih, Eds.\hskip 1em plus 0.5em minus 0.4em\relax Online and Punta Cana, Dominican Republic: Association for Computational Linguistics, Nov. 2021, pp. 8696--8708. [Online]. Available: \url{https://aclanthology.org/2021.emnlp-main.685/}
\BIBentrySTDinterwordspacing

\bibitem{li2023starcoder}
\BIBentryALTinterwordspacing
R.~Li, L.~B. allal, Y.~Zi, N.~Muennighoff, D.~Kocetkov \emph{et~al.}, ``Starcoder: may the source be with you!'' \emph{Transactions on Machine Learning Research}, 2023, reproducibility Certification. [Online]. Available: \url{https://openreview.net/forum?id=KoFOg41haE}
\BIBentrySTDinterwordspacing

\bibitem{roziere2023code}
B.~Roziere, J.~Gehring, F.~Gloeckle, S.~Sootla, I.~Gat, X.~E. Tan, Y.~Adi, J.~Liu, R.~Sauvestre, T.~Remez \emph{et~al.}, ``Code llama: Open foundation models for code,'' \emph{arXiv preprint arXiv:2308.12950}, 2023.

\bibitem{assadi2026maeb}
A.~E. Assadi, I.~Chung, C.~Xiao, R.~Solomatin, A.~Jha, R.~Chand, S.~Singh, K.~Wang, A.~S. Khan, M.~M. Nasser \emph{et~al.}, ``Maeb: Massive audio embedding benchmark,'' \emph{arXiv preprint arXiv:2602.16008}, 2026.

\bibitem{zhao2023get}
Y.~Zhao, L.~Gong, H.~Zhang, Y.~Yu, and Z.~Huang, ``How to get better embeddings with code pre-trained models? an empirical study,'' \emph{arXiv preprint arXiv:2311.08066}, 2023.

\bibitem{huang-etal-2021-cosqa}
\BIBentryALTinterwordspacing
J.~Huang, D.~Tang, L.~Shou, M.~Gong, K.~Xu, D.~Jiang, M.~Zhou, and N.~Duan, ``{C}o{SQA}: 20,000+ web queries for code search and question answering,'' in \emph{Proceedings of the 59th Annual Meeting of the Association for Computational Linguistics and the 11th International Joint Conference on Natural Language Processing (Volume 1: Long Papers)}, C.~Zong, F.~Xia, W.~Li, and R.~Navigli, Eds.\hskip 1em plus 0.5em minus 0.4em\relax Online: Association for Computational Linguistics, Aug. 2021, pp. 5690--5700. [Online]. Available: \url{https://aclanthology.org/2021.acl-long.442/}
\BIBentrySTDinterwordspacing

\bibitem{gong2024cosqa+}
J.~Gong, Y.~Wu, L.~Liang, Z.~Zheng, and Y.~Wang, ``Cosqa+: Enhancing code search dataset with matching code,'' \emph{arXiv preprint arXiv:2406.11589}, 2024.

\bibitem{li2024coircomprehensivebenchmarkcode}
X.~Li, K.~Dong, Y.~Q. Lee, W.~Xia, H.~Zhang, X.~Dai, Y.~Wang, and R.~Tang, ``Coir: A comprehensive benchmark for code information retrieval models,'' in \emph{Proceedings of the 63rd Annual Meeting of the Association for Computational Linguistics (Volume 1: Long Papers)}, 2025, pp. 22\,074--22\,091.

\bibitem{alain2016understanding}
\BIBentryALTinterwordspacing
G.~Alain and Y.~Bengio, ``Understanding intermediate layers using linear classifier probes,'' in \emph{International Conference on Learning Representations (ICLR)}, 2017. [Online]. Available: \url{https://openreview.net/forum?id=HJ4-rAVtl}
\BIBentrySTDinterwordspacing

\bibitem{hewitt2019designing}
\BIBentryALTinterwordspacing
J.~Hewitt and P.~Liang, ``Designing and interpreting probes with control tasks,'' in \emph{Proceedings of the 2019 Conference on Empirical Methods in Natural Language Processing and the 9th International Joint Conference on Natural Language Processing (EMNLP-IJCNLP)}, K.~Inui, J.~Jiang, V.~Ng, and X.~Wan, Eds.\hskip 1em plus 0.5em minus 0.4em\relax Hong Kong, China: Association for Computational Linguistics, Nov. 2019, pp. 2733--2743. [Online]. Available: \url{https://aclanthology.org/D19-1275/}
\BIBentrySTDinterwordspacing

\bibitem{ribeiro2016should}
M.~T. Ribeiro, S.~Singh, and C.~Guestrin, ``" why should i trust you?" explaining the predictions of any classifier,'' in \emph{Proceedings of the 22nd ACM SIGKDD international conference on knowledge discovery and data mining}, 2016, pp. 1135--1144.

\bibitem{lundberg2017unified}
S.~M. Lundberg and S.-I. Lee, ``A unified approach to interpreting model predictions,'' \emph{Advances in neural information processing systems}, vol.~30, 2017.

\bibitem{anselmi2016invariance}
F.~Anselmi, L.~Rosasco, and T.~Poggio, ``On invariance and selectivity in representation learning,'' \emph{Information and Inference: A Journal of the IMA}, vol.~5, no.~2, pp. 134--158, 2016.

\bibitem{lenc2015understanding}
K.~Lenc and A.~Vedaldi, ``Understanding image representations by measuring their equivariance and equivalence,'' in \emph{Proceedings of the IEEE conference on computer vision and pattern recognition}, 2015, pp. 991--999.

\bibitem{penedo2025codeforces}
G.~Penedo, A.~Lozhkov, H.~Kydlíček, L.~B. Allal, E.~Beeching, A.~P. Lajarín, Q.~Gallouédec, N.~Habib, L.~Tunstall, and L.~von Werra, ``Codeforces,'' \url{https://huggingface.co/datasets/open-r1/codeforces}, 2025.

\bibitem{10.1145/3472410.3472445}
\BIBentryALTinterwordspacing
J.~Petrik and D.~Chuda, ``The effect of time drift in source code authorship attribution: Time drifting in source code - stylochronometry,'' in \emph{Proceedings of the 22nd International Conference on Computer Systems and Technologies}, ser. CompSysTech '21.\hskip 1em plus 0.5em minus 0.4em\relax New York, NY, USA: Association for Computing Machinery, 2021, p. 87–92. [Online]. Available: \url{https://doi.org/10.1145/3472410.3472445}
\BIBentrySTDinterwordspacing

\bibitem{zheng2023codegeex}
Q.~Zheng, X.~Xia, X.~Zou, Y.~Dong, S.~Wang, Y.~Xue, Z.~Wang, L.~Shen, A.~Wang, Y.~Li, T.~Su, Z.~Yang, and J.~Tang, ``Codegeex: A pre-trained model for code generation with multilingual benchmarking on humaneval-x,'' in \emph{Proceedings of the 29th ACM SIGKDD Conference on Knowledge Discovery and Data Mining}, 2023, pp. 5673--5684.

\bibitem{ratcliff1988gestalt}
J.~W. Ratcliff and D.~E. Metzener, ``Pattern matching: The gestalt approach,'' \emph{Dr. Dobb's Journal}, vol.~13, no.~7, p.~46, 1988.

\bibitem{voyageai2024voyagecode3}
\BIBentryALTinterwordspacing
{Voyage AI}. (2024, Dec.) {voyage-code-3: more accurate code retrieval with lower dimensional, quantized embeddings}. Voyage AI Blog, accessed March 26, 2026. [Online]. Available: \url{https://blog.voyageai.com/2024/12/04/voyage-code-3/}
\BIBentrySTDinterwordspacing

\bibitem{kryvosheieva2025efficientcodeembeddingscode}
D.~Kryvosheieva, S.~Sturua, M.~G{\"u}nther, S.~Martens, and H.~Xiao, ``Efficient code embeddings from code generation models,'' in \emph{NeurIPS 2025 Fourth Workshop on Deep Learning for Code}.

\bibitem{openai2022textembeddingada002}
\BIBentryALTinterwordspacing
{OpenAI}. (2022, Dec.) New and improved embedding model. [Online]. Available: \url{https://openai.com/index/new-and-improved-embedding-model/}
\BIBentrySTDinterwordspacing

\bibitem{bytedance2024doubaoembeddinglarge}
\BIBentryALTinterwordspacing
{ByteDance Volcano Engine}. (2024) {doubao-embedding-large}. Volcano Engine documentation, accessed March 26, 2026. [Online]. Available: \url{https://www.volcengine.com/docs/84313/1254568}
\BIBentrySTDinterwordspacing

\bibitem{openai2024textembedding3}
\BIBentryALTinterwordspacing
{OpenAI}. (2024, Jan.) New embedding models and api updates. [Online]. Available: \url{https://openai.com/index/new-embedding-models-and-api-updates/}
\BIBentrySTDinterwordspacing

\bibitem{babakhin2025llamaembednemotron8buniversaltextembedding}
\BIBentryALTinterwordspacing
Y.~Babakhin, R.~Osmulski, R.~Ak, G.~Moreira, M.~Xu, B.~Schifferer, B.~Liu, and E.~Oldridge, ``Llama-embed-nemotron-8b: A universal text embedding model for multilingual and cross-lingual tasks,'' 2025. [Online]. Available: \url{https://arxiv.org/abs/2511.07025}
\BIBentrySTDinterwordspacing

\bibitem{zhang2025qwen3embeddingadvancingtext}
\BIBentryALTinterwordspacing
Y.~Zhang, M.~Li, D.~Long, X.~Zhang, H.~Lin, B.~Yang, P.~Xie, A.~Yang, D.~Liu, J.~Lin, F.~Huang, and J.~Zhou, ``Qwen3 embedding: Advancing text embedding and reranking through foundation models,'' 2025. [Online]. Available: \url{https://arxiv.org/abs/2506.05176}
\BIBentrySTDinterwordspacing

\end{thebibliography}

\end{document}